\documentclass[fleqn,10pt]{wlscirep}
\usepackage{amsmath,amssymb}
\title{Staircase Quantum Dots Configuration in Nanowires for Optimized
  Thermoelectric Power}

\author[1,*]{Lijie Li}
\author[2,+]{Jian-Hua Jiang}
\affil[1]{Multidisciplinary Nanotechnology Centre, College of Engineering, Swansea University, Bay Campus, Swansea, SA1 8QQ, UK}
\affil[2]{College of Physics, Optoelectronics and Energy, \& Collaborative Innovation Center of Suzhou Nano Science and Technology, Soochow University, 1 Shizi Street, Suzhou 215006, China}

\affil[*]{L.Li@swansea.ac.uk}

\affil[+]{jianhuajiang@suda.edu.cn}

\begin{abstract}
The performance of thermoelectric energy harvesters can be
  improved by nanostructures that exploit inelastic transport
  processes. One prototype is the three-terminal hopping
  thermoelectric device where electron hopping between quantum-dots
  are driven by hot phonons. Such three-terminal hopping
  thermoelectric devices have potential in achieving high efficiency
  or power via inelastic transport and without relying on
  heavy-elements or toxic compounds. We show in this work how 
  output power of the device can be optimized via tuning the number
  and energy configuration of the quantum-dots embedded in parallel
  nanowires. We find that the staircase energy configuration with
  constant energy-step can improve the power factor over a serial
  connection of a single pair of quantum-dots. Moreover, for a fixed
  energy-step, there is an optimal length for the nanowire. Similarly
  for a fixed number of quantum-dots there is an optimal energy-step
  for the output power. Our results are important for future
  developments of high-performance nanostructured thermoelectric devices. 

\end{abstract}


\begin{document}

\flushbottom
\maketitle

\thispagestyle{empty}

\section*{Introduction}

Thermoelectric energy harvesting has the reverse effect as opposed to
the thermoelectric refrigerator\cite{ed,cam}, and has been studied
extensively in recent decades\cite{rev1,rev2,rev3,ms}. Right from the
invention of the Seebeck and Peltier effects up to now, people have
been using doped semiconductor materials with the aim of increasing
the electrical conductivity and reducing thermal conductivity for a
higher figure of merit. Figures. 1a and 1b show conventional
two-terminal Seebeck thermoelectric energy harvester in its normal and
unfolded geometries, respectively. The configuration in Figure 1b is
very similar to $p$-$n$ junction for solar cells. However, the
metallic contacts in the middle remove the junction barrier and
enables elastic thermoelectric transport. Although Figure 1b reveals
the similarity and difference between a thermoelectric engine and a
solar cell, the mechanism that accounts for the significant difference
of the two devices in efficiency (i.e., solar cells have much higher
efficiency compared with thermoelectric engines) has not been
uncovered. It was found only recently that a $p$-$n$ junction
thermoelectric engine based on hot-phonon-assisted interband
transition can have considerably augmented thermoelectric efficiency
and output power compared to conventional two-terminal thermoelectric
devices with the same material. The intrinsic mechanism
that distinguishes thermoelectric engines and solar cells in their
efficiency and output power is then revealed as akin to inelastic
transport processes in a three-terminal geometry\cite{pn,ourrev} (Figure 1c). 

Other prototypes of three-terminal inelastic thermoelectric devices
include Coulomb coupled quantum-dots (QDs)\cite{rafa,rafael-rev},
phonon-assisted hopping in QD chains (or localized states in
1D or 2D systems) \cite{jiang1,jiang2,pjl1,jiang3,pjl2}, and inelastic
thermoelectric transport across an electronic cavity promoted by
mismatched resonant tunneling at the two-sides of the cavity
\cite{jordan1,jordan2,3tjap,jordan3}. In those devices thermal energy
from the third, insulating terminal of phonon or electronic bath is
converted to electrical energy between the source and the drain (and
vice versa). It was found that to optimize the performance (efficiency
and output power) the energy of the two QDs has to be above
and below the chemical potential around $3k_BT$,
respectively. Experimental developments on three-terminal inelastic 
thermoelectric devices in mesoscopic systems at low-temperature were
established recently\cite{3texp1,3texp2,3texp3}, attracting more and
more researches in the field\cite{ourrev}.

In this work we focus on phonon-assisted hopping thermoelectric
transport in a three-terminal thermoelectric energy harvester. Our
main concern is to optimize the output power of such a device by
tuning the energy configuration of a chain of QDs embedded in
a nanowire. Such a scheme can be used to form a macroscopic
thermoelectric device, since many parallel nanowires can be assembled
together and the nanoscale thermoelectric engines can be connected in
series. We compare the power factor (density) $P=\sigma S^2$ for these
configurations consisting of many serially connected nano- thermoelectric
engines along the $x$ direction (while in $y$-$z$ directions there are
many parallel nanowires, see Figure 2). Specifically, we focus on two configurations: (1) in each nano-engine there is only a single pair of QDs; (2) in each nano-engine there are a chain of $N$ QDs with staircase configuration of energy
(see Figure 1d). We emphasize that the
density of QDs along $x$ direction is the same for all these
situations and the only difference here is the energy
configuration. We show that the power factor is largest for staircase
energy configuration (the main focus of this paper). Particularly we
study the dependence of power factor on the number $N$ and the 
energy-step $dE$ for each hopping. We find that for a given $dE$ there 
is an optimal number $N$ that maximizes the power factor. Our findings
reveal important information for future design of inelastic
thermoelectric devices.

\section*{THREE-TERMINAL THERMOELECTRIC TRANSPORT FOR NANOWIRE QUANTUM-DOTS}

The three-terminal thermoelectric energy harvesting device reported
here is composed of two electrodes on the left and right sides under
room-temperature environment, and a central region comprising
QDs heated by the external phonon bath (Figure 1d). By
absorbing the phonon energy electrons hop from one QD to
another which leads to electrical current against the voltage
gradient. These are the processes that convert thermal energy from
phonon bath to electrical energy\cite{pn}. For realistic devices
operating at room-temperature and above, electron hopping is
efficiently assisted by scattering with optical phonons. Optical
phonon scattering transfer a considerable amount of energy, ranging
from 10~meV to 120~meV for various materials\cite{madelung}. In
addition, electron--optical-phonon scattering time can be as short as
0.1~ps, leading to collision broadening as large as
10~meV\cite{linewidth}. These features make optical-phonon-assisted
hopping thermoelectric transport as promising mechanism for powerful
and efficient thermoelectric energy conversion.

We shall consider many serially connected nano- thermoelectric
engines along the $x$ direction, while in $y$-$z$ directions there are
many parallel nanowires, as shown in Figure 2. Specifically, each
QD is of length $l_{qd}=6$~nm along $x$ direction. The
distance between adjacent QDs is $d=6$~nm. The probability of
finding an electron outside the QD decays exponentially with
the distance away from it with a characteristic length $\xi=2$~nm. A
simplified treatment based on Fermi golden rule yields
the following hopping transition rate from QD $i$ to QD $j$,
\begin{equation}
\Gamma_{i\to j} = 2 \alpha_{ep} \exp( - |x_i-x_j|/\xi ) f_i (1-f_j) N_p(E_i-E_j),
\end{equation}
where the factor of two comes from spin-degeneracy,
$\alpha_{ep}=10$~meV characterizes the strength of electron-phonon
scattering, $f_i$ and $f_j$ are the probability of finding electron on
QDs $i$ and $j$, respectively. The $x$ coordinates of the two QDs are
$x_i$ and $x_j$, respectively, while their energies are $E_i$ and
$E_j$, respectively. Here the phonon distribution function is given by
\begin{equation}
N_p = \frac{1}{\exp(\frac{|E_i-E_j|}{k_BT_p}) - 1} + \frac{1}{2} + \frac{1}{2}{\rm
  sgn}(E_i-E_j) .
\end{equation}
In our thermoelectric energy harvester, the phonon bath has
temperature higher than the electrodes, i.e., $T_p>T$. The heat from
the phonon bath is then converted into electricity. The electron
distribution in each QD can be described by a Fermi distribution 
\begin{equation}
f_i=\frac{1}{\exp(\frac{E_i-\mu_i}{k_BT}) + 1} .
\end{equation}

From the above, the electric current flowing from QD $i$ to QD $j$ is
given by 
\begin{equation}
I_{i\to j} = e(\Gamma_{i\to j} - \Gamma_{j\to i}) ,
\end{equation}
with $e$ being the charge of a single electron. The linear conductance
of electric conduction between QDs $i$ and $j$ is given by
\begin{align}
G_{ij} & = \frac{e^2}{k_BT} \Gamma^0_{ij} ,\\
\Gamma^0_{ij} & = 2 \alpha_{ep} \exp( - |x_i-x_j|/\xi ) f_i^0 (1-f_j^0)
N_p^0(E_i-E_j) ,
\end{align}
where $\Gamma^0_{ij}$ is the transition rate at equilibrium and the
superscripts $0$ denote equilibrium distributions. Hence each pair of
QDs form a resistor with conductance $G_{ij}$. Hopping conduction is
mapped to conduction in network of resistance (i.e., the
Miller-Abrahams network\cite{ma}). Such method is generalized to
three-terminal hopping conduction in Refs.~\cite{jiang1,jiang2}.
Thermoelectric transport through the system is calculated by solving
the Kirchhoff current equation, i.e., the total current flowing into
QD $i$ is equal to the total current flowing out of QD $j$ for the
Miller-Abrahams network\cite{jiang2}. In this fashion the
electrochemical potentials at each QD. i.e., $\mu_i$'s, are determined
numerically via the method presented in Ref.~\cite{jiang2}.

We shall consider a chain of $N$ QDs with staircase energy
configuration. Each energy step is $dE=E_{i+1}-E_{i}$ (we focus on the
situation with $dE>0$). The total energy difference is $\Delta
E=(N-1)dE$. The first QD has energy $E_1=-\Delta E/2$, while the last
QD has energy $E_N=\Delta E/2$. Here the energy we referred to is the
energy of the lowest two degenerate electronic levels (i.e., spin-up and
spin-down) of the QD. Higher levels in the QDs are ignored due to
their much higher energies as we consider small QDs here. $N=2$ is the
case with a single pair of QDs in a nano-thermoelectric engine.
The energy configuration is chosen to have particle-hole symmetry,
which has been proven to be best for thermoelectric performance as
shown in Refs.~\cite{jordan1,3tjap}.

It is noted that differing from variable range hopping between
randomly localized states in nanowires or higher dimensional systems,
the staircase energy configuration always favours the nearest neighbour
hopping. This is because hopping to farther neighbour QDs
costs larger energy gap and longer distance simultaneously. In
contrast, in variable range hopping, the nearby neighbours may have
larger energy differences compared to farther localized
states. Optimization of the hopping distance in 1D localized system
leads to the Mott’s law\cite{mott} in non-interacting electron systems
(with slight modifications\cite{1dvrh}).

It is necessary to mention that the Fermi golden rule requests that
the energy difference $dE$ between the two electronic states must be
the same as the optical phonon energy (i.e., microscopic energy
conservation). We are interested only in the range with $dE\in (10,
120)$~meV, which can be realized in III-V, II-VI, VI
semiconductors. Considering the electron--optical-phonon scattering
rate for most of those semiconductors are around 0.1~ps. Our model
calculation hence captures the main physics of the system. We mention
that acoustic-phonon scattering near the Debye frequency is also very
efficient. In our calculation, modifying $dE$ may need to be fulfilled
by changing the materials for nanowire-QDs. Nevertheless, our study
reveals for a given $dE$ (i.e., a given material) the number of QDs in
a single nanowire that optimizes the power factor, as well as how
such an optimal number varies with $dE$. These information are useful for
future material design of nanowire-QD thermoelectric devices.

Beside the inelastic hopping conduction, there is also elastic
transport through the system. The elastic transport defines pure
electron quantum tunneling mechanism between QDs and electrodes. A
resonant tunneling mechanism is exploited to describe such conduction 
process. Note that since we consider QDs with considerably
large energy differences (much larger than coupling between quantum
dots) sequential tunneling between QDs is suppressed. The
dominant contribution comes from single QD resonant tunneling\cite{jiang2},
where each QD forms one of such resonant tunneling conduction channel
independently. Hence the elastic conduction contributes
to the electric current via

\begin{align}
I_{el} &= G_{el} V , \quad G_{el} = \sum_i G_i ,\\
G_i&=\frac{2e^2}{h} \int \frac{d\varepsilon}{k_BT} \frac{
  \gamma_{Li}\gamma_{Ri}}{(\varepsilon - E_i)^2 +
  (\gamma_{Li}+\gamma_{Ri})^2/4} f^0(\varepsilon)[1-f^0(\varepsilon)] .
\end{align}
Here $V$ is the voltage across the source and drain electrodes, $h$ is
the Planck constant, $G_{el}$ denotes the elastic conductance,
$f^0(\varepsilon)=1/[\exp(\varepsilon/(k_BT))+1]$ (we set the
electrochemical potential at equilibrium as energy zero). The tunnel
coupling between the QD $i$ and the left (right) electrode is
$\gamma_{Li}$ ($\gamma_{Ri}$). We shall set the coordinate of the left
electrode as $x=0$, while the right electrode has $x=L_{tot}$ with
$L_{tot}=Nl_{qd}+(N-1)d+2l_b$ where $l_b$ is the distance between the first
(last) QD and the left (right) electrode. The tunnel coupling is hence
$\gamma_{Li}=t_0\exp(-x_i/\xi)$ and
$\gamma_{Ri}=t_0\exp(-(L_{tot}-x_i-l_{qd})/\xi)$ with $t_0=100$~meV that
characterizes hybridization energy of closely coupled QDs.
We emphasize that the elastic current $I_{elas}$ does not vary with
the temperature of the phonon bath $T_p$ since it originates purely from
the quantum tunneling instead of coupling with phonons. In fact, in
our thermoelectric engine, elastic conduction dissipates the electric
energy generated by inelastic hopping into Joule heating.

Thermoelectric transport in our system the linear-response regime can
be described by the coupled electric and heat conduction equation 
\begin{align}
\left( \begin{array}{c} I_e\\ I_Q \end{array}\right) =
\left( \begin{array}{cccc}
G & L \\
L & K \\
\end{array}\right) \left(\begin{array}{c}
V \\ \frac{T_p-T}{T} \end{array}\right) \ ,\label{FIRO}
\end{align}
where $G=G_{in}+G_{el}$ with $G_{in}$ being the inelastic
conductance. It was shown in Ref.~\cite{jiang2} that
{$L = G_{in} (\overline{E_R}-\overline{E_L})/e$ where $\overline{E_R}$ ($\overline{E_L}$)}
is the average energy of electrons entering into the right (left)
electrode. For instance, hopping thermoelectric transport in a single
pair of QDs gives $L= G_{in} (E_2-E_1)/e$. Hopping for a chain of $N$
QDs with staircase energy configuration yields $L= G_{in}
(E_N-E_1)/e$. We emphasize that elastic tunneling does {\em not}
contribute to the Seebeck effect here, which is the essential
difference between three-terminal and conventional
thermoelectric effects. The Seebeck coefficient for
the phonon-driven three-terminal thermoelectric effect is then
\begin{equation}
S \equiv \frac{L}{TG} = \frac{k_B}{e} \frac{G_{in}
  (N-1)dE}{(G_{in}+G_{el})k_BT} . \label{seebeck}
\end{equation}

For our system to work as an thermoelectric energy harvester, the
inelastic conduction should dominate over the elastic conduction.
The inelastic, elastic and total conductivity are plotted in Figure 3
(a) for a nano thermoelectric harvester with a single pair of QDs as
functions of energy step $dE$ for $dE\in (10, 120)$~meV. Both
the inelastic and elastic conductivity decreases with increasing
$dE$. The elastic conductivity is reduced as the first QD has lower
energy below the electrochemical potential while the second QD is
higher above the electrochemical potential, leading to less effective
conduction. The inelastic conductivity is also reduced due to the larger
thermal activation energy $dE$ and exponentially decreased phonon
number $N_p^0$. Therefore at very large energy difference $dE$ the
elastic conductivity may be more important. In reality, the elastic
conduction also dominates in the small $dE$ regime, which we ignored
in this study. For a chain of QDs with $N=10$, the results in Figure 3(b) shows that
the elastic conduction is much reduced, since tunneling over a
longer distance is exponentially suppressed. The conductivity is then
dominated by inelastic hopping in long chains of QDs. 

Next we examine the conductivity as a function of the length of the
chain. We show the results in Figures 3(c) and 3(d) for $dE=10$~meV
and 30~meV, respectively. As the number of QDs increases both the
inelastic hopping conductivity and elastic tunneling conductivity
decreases. However, the elastic conductivity decreases much
rapidly. The initial decay of hopping conductivity is sub-exponential,
since increase the number of hopping is similar to increase the number
of resistors. However, for large $N$, as the energy of the first (last)
few QDs is much lower (higher) than the electrochemical potential, the
hopping rates are suppressed by the exponentially small availability
(occupation) of the final (initial) state. The decrease of
conductivity at large $N$ is hence exponential. Such exponential
decrease become stronger for larger $dE=30$~meV as shown in Figure
3(d).

\section*{POWER FACTOR FOR DIFFERENT ENERGY CONFIGURATIONS}
We then study the power factor $P=\sigma S^2$ for various energy step
$dE$ and length of the QDs chain $N$. We remark again that for all
situations the density of QDs is the {\em same}, according to our
geometry of the nanowire QDs. The conductivity is calculated via
$\sigma = G l / A$ where $l$ and $A$ are the length and area of a
single nano thermoelectric engine. Here the area is determined by the
density of nanowires as $A^{-1}=10^{15}$~m$^{-2}$\cite{nw1,nw2}. By
focusing on the scale independent conductivity $\sigma$ and power
factor $\sigma S^2$ we are able to discuss ways of optimizing the power
factor by engineering each nano thermoelectric element. In this way, 
the variation of the power factor $\sigma S^2$ is a sole consequence
of the energy configuration (rather than geometry) in each nano
thermoelectric engine.

In Figure 4(a) we show the dependences of power factor $\sigma S^2$
and the Seebeck coefficient $S$ on the energy step $dE$ for a nano
device with a single pair of QDs. It is found that the Seebeck
coefficient $S$ increases monotonically with the energy step $dE$,
which is consistent with Eq.~(\ref{seebeck}). As a consequence of
competition between the conductivity and the Seebeck coefficient, the
power factor is optimized around $dE=3k_BT$. For a chain of $N=10$
QDs, the power factor is maximized at a much lower $dE$, as shown in
Figure 4(b). This is due to the more rapid decay of the conductivity
as shown in Figure 3(b).

Similarly, the dependence of the number of QDs $N$ for a given energy
difference $dE$ also has a peak, as shown in Figures 4(c) and
4(d). In Figure 4(c) we plot the power factor $\sigma S^2$ and the
conductivity $\sigma$ as functions of the number of QDs $N$ in a
single nano device for $dE=10$~meV. The power factor is maximized at
$N=21$. This maximum also appears as a consequence of the competition
of the conductivity and the Seebeck coefficient when the number $N$ is
increased. The Seebeck coefficient increases as the total energy
difference $\Delta E=(N-1)dE$ increases, while the conductivity decays
exponentially with the number of QDs for large $N$. Since such
exponential decay of conductivity is more severe for larger energy
step $dE$, the maximum appears at a smaller number of QDs $N$ for
$dE=30$~meV, as shown in Figure 4(d). 

To have a global view of the dependence of the power factor on the
energy configuration of QDs, we plot the $\sigma S^2$ for various $N$
and $dE$ in Figure 5. It is seen that for each $dE$ there is an
optimized $N$ at which the power factor is maximized. For smaller $dE$
the optimal $N$ is larger. More importantly, the maximal power factor
is greater. Our study thus reveal the optimal energy configurations
for powerful three-terminal thermoelectric energy harvester. In
reality, it is important to find the energy $dE$ that optimize the
phonon-assisted hopping rate and the conductivity. Fixing such a $dE$
one can find an optimal number of QDs $N$ that form the maximal output
power for a given material.

\section*{CONCLUSION}
We study the optimization of energy configurations of thermoelectric
energy harvester assembled by many nano thermoelectric elements. Each
nano device contains $N$ QDs of staircase energy configuration with
energy step $dE$. It is found that such energy configuration is {\em
  better} than the situation studied before: each nano thermoelectric
element contains only $N=2$ QDs. More importantly, we find that for
each given energy step $dE$ there is an optimal number of QDs $N$ that
maximizes the power factor. Such optimization yields higher output
power when $dE$ is smaller. Finally, we argue that our design is also
better in thermoelectric power factor than hopping in a chain of QDs
with random energy configuration. This is because the conductivity of
such a random energy QDs chain is lower than that of the nano device
with a single pair of QDs (when its parameters are optimized). On the
other hand, the Seebeck coefficient is fluctuating around
zero\cite{jiang2}, yielding relatively low Seebeck
coefficient. Therefore, the power factor can be relatively lower as
compared to the situation of assembled thermoelectric energy harvester
with each nano-scale element contains a single pair of QDs. Our design
of staircase energy configuration can have much better output power
than the sigle-pair of QDs nano device. Therefore, our study is 
valuable for future design of nanostructured thermoelectric devices. Future study should also include the effect of parasitic heat conduction due to, e.g., phonon thermal conductivity, which may reduce the figure of merit, although it is usually much smaller in nanowires than in bulk materials.

\section*{METHODS}
The power factor is calculated by computing the conductivity $\sigma$
and Seebeck coefficient $S$. From Eq.~(\ref{seebeck}), the essential
quantities of interest are the conductivity for both inelastic and
elastic transport processes. The conductivity for elastic processes
are calculated via Eqs.~(7) and (8). The hopping conductivity is
calculated via solving the Kirchhoff current equation for
Miller-Abrahams resistor network numerically. The key quantities to be
calculated are the ``local voltage'' $V_i=\mu_i/e$ for all $i$
labeling the QDs. According to Ref.~\cite{jiang2} the equations to be
solved are
\begin{align}
\sum_j A_{ij}V_j = z_i
\end{align}
where
\begin{align}
  A_{ii} &= G_{iL}+G_{iR} + \sum_{k\ne i}G_{ik}, \quad\quad  A_{ij} = -
  G_{ij} \quad ({\rm for}\ i\ne j), \quad\quad z_i = G_{iL} V_L + G_{iR} V_R 
\end{align}
The left electrode has voltage $V_L=V/2$, while the
right electrode has voltage $V_R=-V/2$. The conductance $G_{iL}$
($G_{iR}$) is finite only when $i$ labels the first (last) QD, and
$G_{1L}=G_{NR}=\frac{2e^2}{k_BT\hbar}t_0\exp(-l_b/\xi)$. The hopping
conductance between QDs are given by Eqs.~(5) and (6). By setting
$V=0.01$ and solving the above equation, we obtain $V_i$ for all $i$.
The electrical current flowing through the system due to hopping is
then calculated via $I_e^{in}=G_{1L}(V_L-V_1)$ using the numerically
obtained $V_1$. The inelastic conductance of the nanowire-QDs system
is then $G_{in}=I_e^{in}/V$.

\bibliography{paper}

\section*{Acknowledgements}

LL appreciates the support of college of engineering, Swansea
University. JJ would like to thank the faculty start-up funding of
Soochow University for support. All authors are grateful for the
encouragement from Prof. Yoseph Imry at Weizmann Institute of 
Science, Israel. 

\section*{Author contributions statement}

All authors contributed equally to shaping the idea, finding the appropriate techniques, and writing the paper.

\section*{Additional information}

The authors declare no competing financial interests.

\newpage

\begin{figure}
\centerline{\includegraphics[height=10.cm]{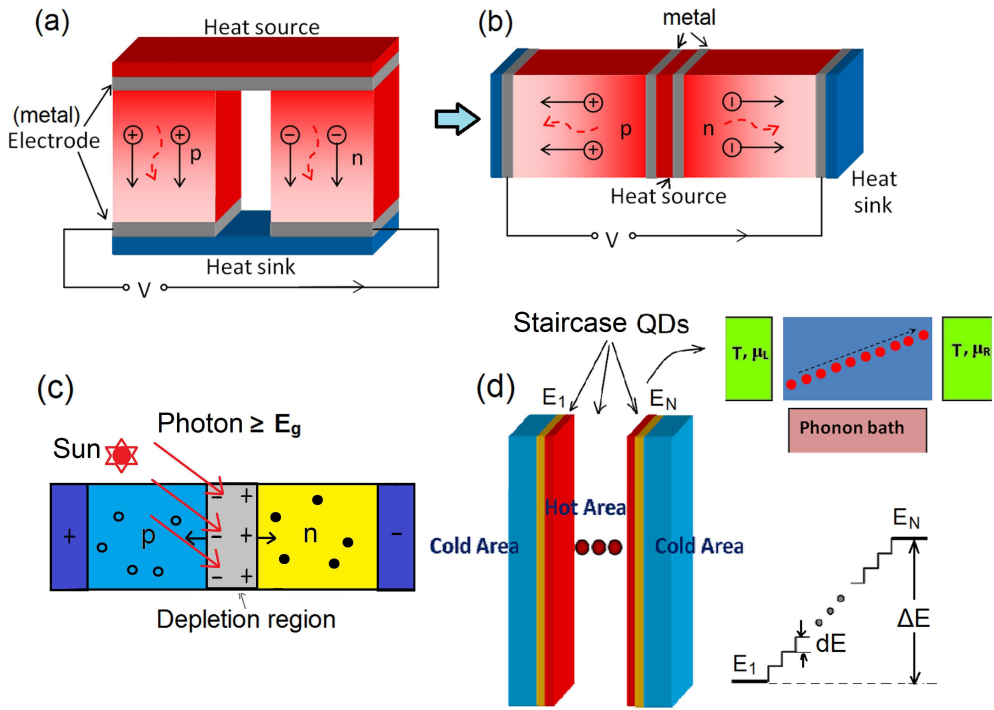}}
\caption{(a) Schematic of the conventional thermoelectric
    energy harvester that converts heat to electricity. (b) Unfolded
    geometry of the thermoelectric energy harvester. (c) Working
    principle of a solar cell, a three-terminal device akin
    to inelastic (i.e., photon absorbing) processes. (d) Schematic of
    staircase quantum-dots thermoelectric harvester. Heat from hot
    phonon bath is exploited to generate electricity via
    phonon-assisted electron hopping in a chain of quantum-dots with
    staircase energy configuration. The energy diagram for this device
    is illustrated as well. Each energy step is $dE$. For a chain of
    $N$ quantum-dots the total energy difference is $\Delta E=(N-1)dE$.}
\label{fig_1}
\end{figure}

\begin{figure}
\centerline{\includegraphics[height=5.cm]{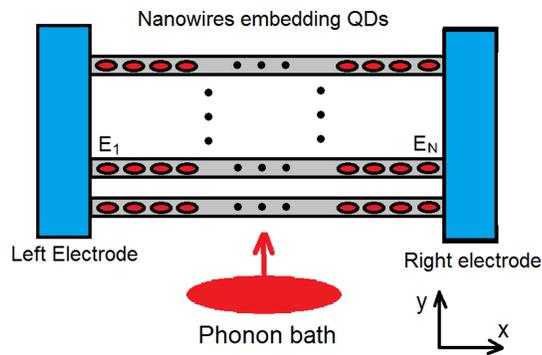}}
  \caption{ Schematic diagram of the thermoelectric energy
    harvester based on series of quantum dots embedded in parallel
    nanowires.} 
    \label{fig_2}
\end{figure}

\begin{figure}
\centerline{\includegraphics[height=9.cm]{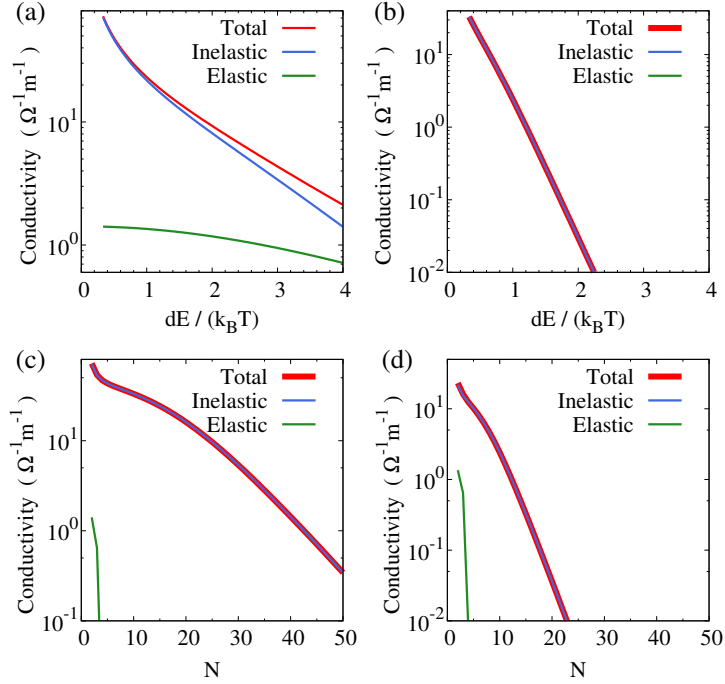}}
   \caption{ (a) and (b) : Inelastic, elastic, and total
    conductivity as functions of $dE$ for (a) a single pair of QDs and
    (b) a chain of $N=10$ QDs with staircase energy configuration. The
    range of $dE$ is between 10 and 
    120~meV. $k_BT=30$~meV. (c) and (d): Inelastic, elastic, and total
    conductivity as functions of the number of QDs in a single
    nano-device with staircase energy configuration for (c)
    $dE=10$~meV and (d) $dE=30$~meV. }
     \label{fig_3}
\end{figure}

\begin{figure}
\centerline{\includegraphics[height=9.cm]{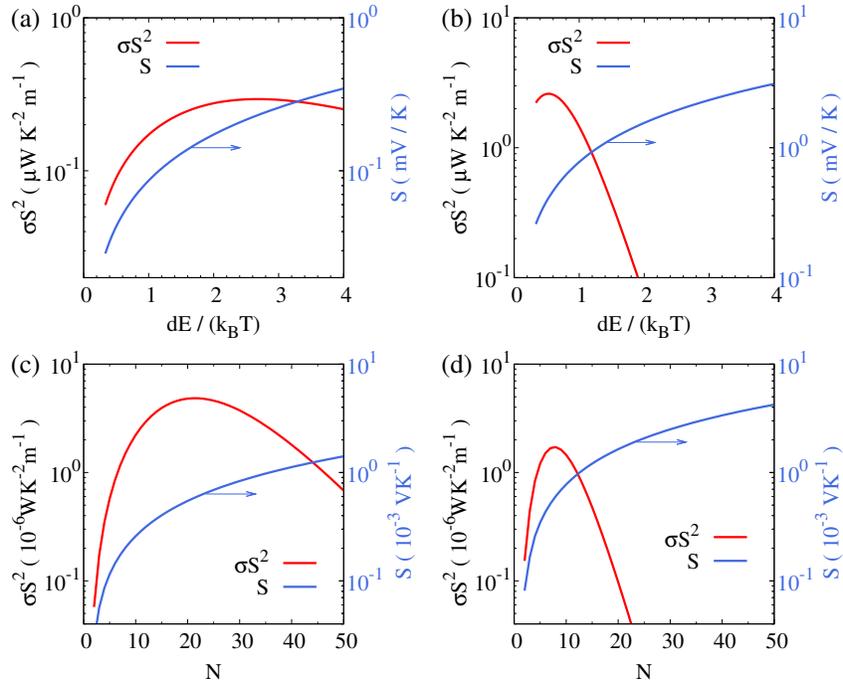}}
\caption{(a) and (b): Power factor $P=\sigma S^2$ and
    Seebeck coefficient $S$ as functions of energy step $dE$ for (a) a
    single pair of QDs and (b) a chain of $N=10$ QDs with staircase
    energy configuration. (c) and (d): Power factor $P=\sigma S^2$ and
    Seebeck coefficient $S$ as functions of the number of QDs in a single
    nano-device with staircase energy configuration for (c)
    $dE=10$~meV and (d) $dE=30$~meV. $k_BT=30$~meV. }
     \label{fig_4}
\end{figure}

\begin{figure}
\centerline{\includegraphics[height=8cm]{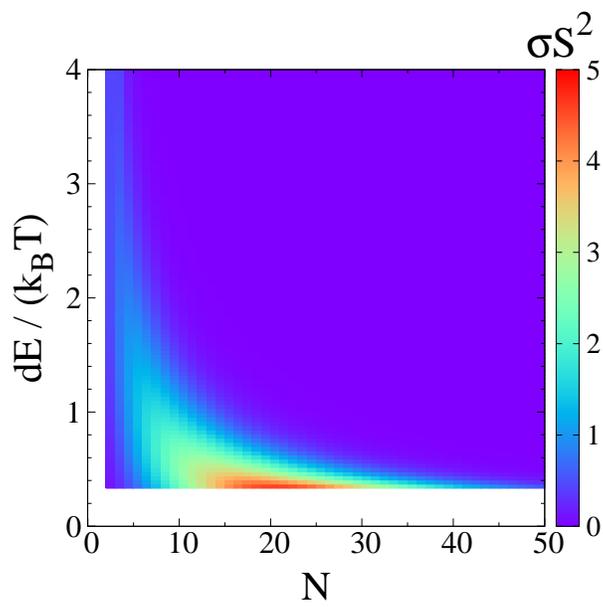}}
 \caption{Power factor $P=\sigma S^2$ as a function of the
    energy step $dE$ and the number of QDs in a single
    nano-device with staircase energy
    configuration. $k_BT=30$~meV. Other parameters are specified in
    the main text.}
 \label{fig_5}   
\end{figure}

\end{document}